\documentclass[%
 aip,
 twocolumn, 
 amsmath,amssymb,
 reprint,%
]{revtex4-2}

\usepackage{graphicx}
\usepackage{dcolumn}
\usepackage{bm}
\usepackage{soul} \usepackage{xcolor} 
\renewcommand\hl{} 

\newif\ifsec\sectrue
\begin{document}

\preprint{AIP/123-QED}

\title
{Hole concentrations in doped gray $\alpha$-Sn on InSb and CdTe measured with infrared ellipsometry}

\author{Jaden R.\ Love}
\author{Carlos A.\ Armenta}
\author{Atlantis K.\ Moses}%
\author{Haley B.\ Woolf}
\affiliation{Department of Physics, New Mexico State University, MSC 3D, P.\ O.\ Box 30001, Las Cruces, NM 88003-8001, USA}
\author{Jan Hrabovsk\'y}
\affiliation{Department of Physics, New Mexico State University, MSC 3D, P.\ O.\ Box 30001, Las Cruces, NM 88003-8001, USA}
\affiliation{Charles University, Faculty of Mathematics and Physics, Ke Karlovu 5, 121 16 Prague, Czech Republic}
\author{Stefan Zollner}
 \email{zollner@nmsu.edu}
 \homepage{http://femto.nmsu.edu.}
\affiliation{Department of Physics, New Mexico State University, MSC 3D, P.\ O.\ Box 30001, Las Cruces, NM 88003-8001, USA}

\author{Aaron N.\ Engel}
\affiliation{Materials Department, University of California Santa Barbara, Santa Barbara, California 93106, USA}
\author{Christopher J.\ Palmstr\o{}m}
\affiliation{Materials Department, University of California Santa Barbara, Santa Barbara, California 93106, USA}
\affiliation{Electrical and Computer Engineering Department, University of California Santa Barbara, Santa Barbara, California 93106, USA.}
\date{\today}

\begin{abstract}
Gray tin ($\alpha$-Sn) layers with 30 nm thickness were grown on InSb (001) substrates using molecular beam epitaxy. The surface preparation of the substrates was adjusted to achieve either $n$-type or $p$-type doping in the $\alpha$-Sn layer. Fourier-transform infrared ellipsometry was used to find the temperature-dependent dielectric function of the $\alpha$-Sn layers from 0.03 to 0.8 eV and from 10 to 300 K. Because of the inverted band structure of $\alpha$-Sn, the spectra show a strong absorption peak at 0.45 eV due to transitions from the inverted $\Gamma_7^-$ ``electron'' valence band to the $\Gamma_8^+$ heavy hole valence band. Applying the Thomas-Reiche-Kuhn f-sum rule, the integrated oscillator strength of this peak was used to calculate the heavy hole concentration as a function of temperature. For a nearly intrinsic $\alpha$-Sn layer, the heavy hole concentration agrees well with predictions based on degenerate Fermi-Dirac statistics. Deviations from the intrinsic $\alpha$-Sn carrier concentrations are attributed to substrate surface preparation leading to the diffusion of  donor or acceptor ions into the $\alpha$-Sn layer  causing $n$-type or $p$-type doping. 
\end{abstract}


\maketitle

\ifsec\section{Introduction}\fi \label{intro}

Because of strong relativistic effects in heavy elements (Darwin shift), the band structure of  $\alpha$-Sn is inverted. In comparison to germanium, the $\Gamma_7^-$ ``electron" band moves between the degenerate $\Gamma_8^+$ heavy and light ``hole" bands and the $\Gamma_7^+$ split-off hole band. It thus becomes a valence band with a negative curvature. In the absence of strain, the $\Gamma_8^+$ heavy hole and light ``hole" bands are degenerate. This makes $\alpha$-Sn a zero-gap semimetal. The heavy hole band is curved like a valence band, while the light ``hole" band is a conduction band.\cite{GrPa63,Ew68,LaEw71,GeIv76} See Fig.\ 1 in Ref.\ \citenum{GrPa63} or Fig.\ 1 in Ref.\ \citenum{CaZo19} for a schematic band structure. Epitaxial growth of $\alpha$-Sn on InSb stabilizes the diamond-like crystal structure at room temperature and avoids transformation into the metallic $\beta$-Sn phase.\cite{EnDe24} The small lattice mismatch of $-$0.15\% between $\alpha$-Sn and InSb leads to an antisplitting of the $\Gamma_8^+$ bands equal to 12 meV for growth of $\alpha$-tin on an InSb (001) surface. Technically, $\alpha$-Sn on InSb is a Dirac semimetal.\cite{CaZa18} However, this splitting is small compared to the energy scale investigated here. For our purposes, it is sufficient to treat $\alpha$-Sn on InSb as a topologically trivial zero-gap semimetal. 

This inverted band structure allows optical interband transitions from the $\Gamma_7^-$ ``electron" band to the $\Gamma_8^+$ heavy and light ``hole" bands, which have traditionally been observed using magnetoreflection, magnetotransmission, and thermoreflectance measurements in $\alpha$-Sn and HgTe.\cite{GrPi70,GuRi73,DoMy78,RaVe73} A strong infrared $\bar{E}_0$ peak due to these transitions was recently found at 0.45 eV in Fourier-transform infrared ellipsometry measurements of the $\alpha$-Sn optical constants.\cite{CaZa18} A careful analysis of the band structure of $\alpha$-Sn showed\cite{Zo24} that this infrared peak is mostly due to transitions from $\Gamma_7^-$ into the $\Gamma_8^+$ heavy hole band because of its large effective mass. Transitions into the $\Gamma_8^+$ light ``hole" band are much weaker (because of the smaller effective mass). The $\bar{E}_0$ energy shows the usual decrease\cite{ZoAb22} with increasing temperature for HgTe,\cite{DoMy78,RaVe73} but is nearly \hl{independent} of temperature for $\alpha$-Sn. \cite{GrPi70}

The purpose of this article is to apply the Thomas-Reiche-Kuhn f-sum rule\cite{St63,AlDe72,Sm98} to calculate the heavy hole concentration from the integrated oscillator strength of the infrared $\bar{E}_0$ peak at 0.45 eV. For a nearly intrinsic (undoped) $\alpha$-Sn layer, the resulting heavy hole concentrations from 10 to 300 K are in excellent agreement with a degenerate Fermi-Dirac carrier statistics model, which considers the electrons located at the nearly degenerate $L$-valleys.\cite{Zo24} (The simple two-band approximations given by Ref.\ \citenum{GeIv76} are not accurate because of the small indirect gap.) Depending on the preparation of the InSb substrate before growth, donor or acceptor doping of $\alpha$-Sn with Sb or In can be achieved. This leads to observable changes in the infrared spectra and the resulting heavy hole concentrations. 

One might also wonder if doping or thermally activated carriers in intrinsic $\alpha$-\hl{Sn} might cause changes to the interband transitions at higher energies, especially the $E_1$ and $E_1+\Delta_1$ critical points.\cite{CaZa18,ViHo85} A detailed study of the temperature dependence of the energies and broadenings of the critical points of intrinsic $\alpha$-Sn was already performed, but no anomalies were found in comparison to Ge.\cite{ViHo85,ViLo84} For very high electron concentrations in the $L$-valley, a Fermi level singularity\cite{XuFe17} and Pauli blocking\cite{ArZa25} of the $E_1$ and $E_1+\Delta_1$ transitions are expected.  These effects become noticeable in Ge at electron concentrations of the order of 10$^{20}$ cm$^{-3}$, about one order of magnitude or more higher than those considered here. Such studies of the interband critical points in the range of 1 to 6 eV require a completely different equipment set and we therefore did not pursue them for our epilayers.

\ifsec\section{Experimental methods}\fi \label{methods} 

\ifsec\subsection{Molecular beam epitaxy and characterization}\fi \label{MBE}

The $\alpha$-Sn films were grown by molecular beam epitaxy on nominally undoped InSb(001) (\hl{Wafer Technology} Ltd.) with an indium rich c(8$\times$2) reconstruction (AE225) and with an antimony rich c(4$\times$4) reconstruction (AE227) of the InSb substrate. The substrates were gallium-bonded to molybdenum substrate holders. 

The InSb(001)-c(8$\times$2) reconstruction is prepared via atomic hydrogen clean (MBE Komponenten GmbH) to remove oxides at 285$^\circ$C followed by a light anneal at 360$^\circ$C to smoothen the surface. The cleaning recipe was periodically verified using scanning tunneling microscopy. Temperature measurements are uncalibrated. The InSb(001)-c(4$\times$4) reconstruction was prepared by annealing the InSb(001)-c(8$\times$2) substrate in a primarily Sb$_4$ overpressure above 430$^\circ$C and then rapidly cooling. For growth on both InSb reconstructions, the substrate was allowed to cool to room temperature prior to initiation of $\alpha$-Sn growth. 

Desired reconstructions were confirmed by reflection high energy electron diffraction (RHEED) immediately prior to the initiation of $\alpha$-Sn growth. The $\alpha$-Sn growth rate was monitored by RHEED oscillations and periodically calibrated by Rutherford backscattering spectrometry. Further details on the growth details and chemistry of the $\alpha$-Sn epilayers can be found in Ref.\ \citenum{EnDe24}. 

High resolution x-ray diffraction (HRXRD) was used to measure $\omega-2\theta$ scans using the Empyrean x-ray diffractometer from Malvern PANalytical.\cite{Malvern} The Empyrean is a five-axis x-ray diffractometer that uses an in-plane vertical goniometer and a horizontal sample stage. X-rays are generated from a copper target using a 45 kV potential. The incident beam optics consists of a two-bounce Ge(220) hybrid monochromator with a 10 mm beam mask and a 1/32$^\circ$ divergence slit. The diffracted beam optics includes a programmable anti-scatter slit (PASS), 0.04 rad Soller slit, 0.02 mm Ni filter, and the PIXcel line detector in receiving slit mode. 

\begin{figure}
\includegraphics[width=\columnwidth]{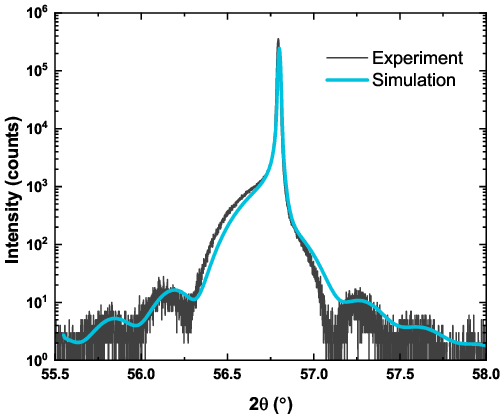}
\caption{High-resolution x-ray diffraction data for sample AE225 compared to a simulation for a 30 nm thick $\alpha$-Sn layer on InSb (001).}
\label{XRD}
\end{figure}

For the analysis of HRXRD data the program Epitaxy SmoothFit,\cite{Malvern} was used to simulate the experimental data based on user input describing the layer composition and lattice constants for the $\alpha$-Sn layer and InSb substrate.\cite{CaZo19} The simulation gives a layer thickness of 29.7 $\pm$ 0.7 nm which is consistent with the nominal thickness, see Fig.\ \ref{XRD}. The 30 nm $\alpha$-Sn layer is assumed to be pseudomorphic on the InSb (100) substrate.\cite{CaZo19} As already mentioned, the pseudomorphic out-of-plane strain in the $\alpha$-Sn layer ($\epsilon_{\perp}$=0.13\%) leads to a very small splitting of the $\Gamma_8^+$ bands, which can be ignored for our mid-infrared energy scales.

\begin{figure*}
\includegraphics[width=\textwidth]{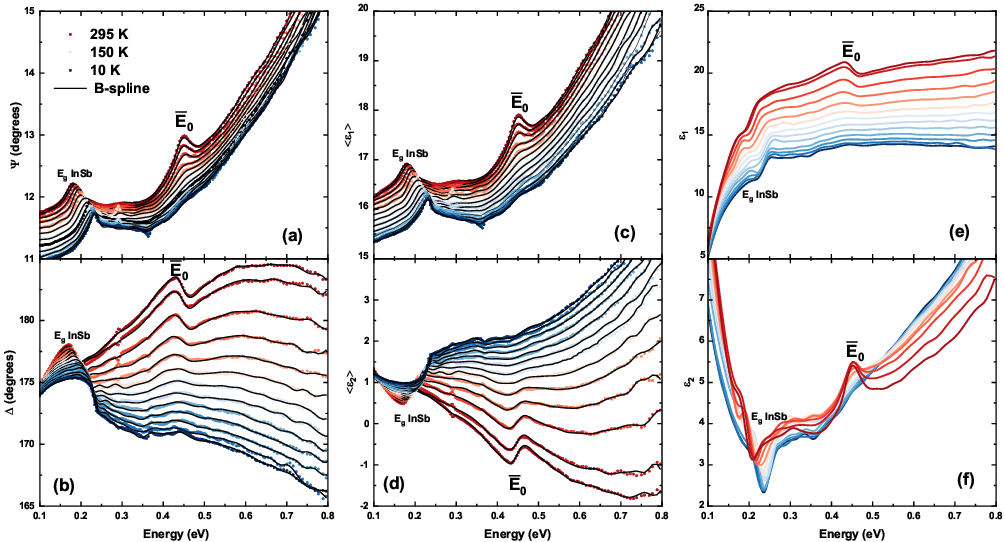}
\caption{Ellipsometric angles $\psi$ (a) and $\Delta$ (b), real (c) and imaginary (d) parts of the pseudodielectric function $\left<\epsilon\right>$ of undoped $\alpha$-Sn on an InSb substrate (sample AE225), and real (e) and imaginary (f) parts of the dielectric function $\epsilon$ of the $\alpha$-Sn layer. The symbols show experimental data, while the lines were obtained from a basis spline fit. The color scale indicates the increase of the sample temperature from 10 K (blue) to 295 K (red)}
\label{psiDel}
\end{figure*}

\ifsec\subsection{Temperature-dependent infrared ellipsometry}\fi \label{FTIR-SE}

The ellipsometric angles $\psi$ and $\Delta$ were measured from 0.05 to 0.8 eV at an incidence angle of 70$^\circ$ on a J.\ A.\ Woollam FTIR-VASE Mark II Fourier transform infrared spectroscopic ellipsometer.\cite{Fu07,ToHi16,Sc04} Since there are no sharp features (such as infrared-active phonons) in the spectra from these samples, a resolution of 64 cm$^{-1}$ was selected to increase the signal to noise ratio. Zone-averaging of the polarizer ($\pm$45$^\circ$) and the analyzer (0$^\circ$, 90$^\circ$) was used to reduce systematic errors. 500 scans were acquired at each of the 15 positions per compensator rotation. 

To obtain temperature-dependent spectra,\cite{ZoAb22} the samples were mounted with silver paint on a copper sample holder in a Lakeshore ST-400 ultrahigh vacuum (UHV) cryostat with diamond windows. The back surface of the InSb substrate was roughened to keep the depolarization due to back side reflections below 1\%. 
The samples were large enough ($>$100 mm$^2$) to avoid depolarization due to light reflected by the copper sample holder. 

The $\alpha$-Sn phase is stabilized at room temperature by the InSb substrate and the phase is sensitive to increasing pressure and temperature. To avoid scratching the surface, the sample backside was fixed to the cold finger stage using silver conductive paint and light pressure was applied using a microfiber cloth to level the sample and maximize contact with the stage. The silver paint cured overnight at room temperature.  The sample temperature is measured from a silicon diode mounted to the sample stage. The UHV cryostat chamber is evacuated for several days prior to low temperature measurements (pressure$<$10$^{-8}$ Torr) to remove water vapor and other contaminants from the chamber. Removing contaminants via baking or cleaning was likely to damage the $\alpha$-Sn and was avoided.  

The cryogen reservoir of the cryostat is connected to a closed loop system that consists of a FA-50L helium compressor (Sumitomo Corporation) and the gas handling system (GHS) with cold head (LakeShore Cryotronics).  The cooling procedure involves three phases: the first is the evacuation of the GHS, the second is room temperature circulation of helium, and the third is the final cool down where the base temperature can be achieved. A 335 Lakeshore temperature controller is linked to the data collection software, WVASE-IR, where an automated series was programmed to run from 300 K down to 10 K in 25 K steps. One scan was taken at each temperature and the series was run over the course of several days. Each scan took 4.5 hours to complete, and the temperature was allowed to stabilize within 0.5 K of the target then holding the target for at least 10 minutes before starting the next scan.  

\ifsec\subsection{Analysis of ellipsometry spectra}\fi \label{analysis} 

The ellipsometric angles\cite{Fu07,ToHi16} $\psi$ and $\Delta$ for a nominally undoped $\alpha$-Sn layer on InSb (sample AE225) are displayed in Fig.\ \ref{psiDel}. $\psi$ shows a monotonic increase with photon energy and two peaks around 0.2 eV (from the band gap of the InSb substrate) and at 0.45 eV (due to the $\bar{E}_0$ intervalence band transition in the $\alpha$-Sn layer). The same two structures also show up as wiggles in the ellipsometric angle $\Delta$. The band gap of the substrate decreases with increasing temperature.\cite{RiAr23} The $\bar{E}_0$ peak of the $\alpha$-Sn layer is quite prominent at room temperature and disappears at lower temperatures. The same experimental data is also displayed as a pseudodielectric function in Fig.\ \ref{psiDel}. Because of interference effects, the absorption peaks in the substrate and in the $\alpha$-Sn layer appear in the real (not imaginary) part of the pseudodielectric function.

Determining the optical constants (dielectric function $\epsilon$) of the $\alpha$-Sn layer requires two pieces of information as input: (1) The thickness of the $\alpha$-Sn layer was found to be 29.7$\pm$0.7 nm using high-resolution (004) x-ray diffraction.\cite{CaZo19} (2) The temperature dependence of the dielectric function of the InSb substrate was obtained previously.\cite{RiAr23} This information allows a direct inversion of the Fresnel equations to find $\epsilon$ of $\alpha$-Sn at each photon energy.\cite{ToHi16} To avoid large point-to-point fluctations, it is more convenient to expand $\epsilon$ using Kramers-Kronig consistent basis spline (B-spline) functions\cite{MoTi19,Li22} with a node spacing of 20~meV. The imaginary part $\epsilon_2$ was forced to be nonnegative. Figure \ref{psiDel} shows the experimental data (symbols) in comparison to the B-spline fit (lines). The same procedure was used at all temperatures and for all samples. This B-spline fit is more predictive and less biased than another commonly used method, in which the dielectric function of the layer is expanded as a series of oscillators.\cite{Fu07} 

Like many digital filters, our B-spline fit introduces oscillations in the data. These oscillations are on the order of 0.1 on the scale of $\epsilon_2$ (in Fig.\ 4) or 5$\times$10$^{17}$ cm$^{-3}$ for the heavy hole concentration (in Fig.\ 5). These oscillations can be minimized through a careful choice of the node spacing of the B-spline fit with a statistical method\cite{Li17} or with a Fourier analysis of the spectra.\cite{LeKi19} We did not do this here and instead used a fixed node spacing of 20~meV, which visually seemed to provide the best results. Our method therefore has much worse sensitivity than typical electrical measurements, which can determine carrier concentrations as low as 10$^{15}$ cm$^{-3}$, see Refs.\ \citenum{HoMe89,OlKa15}. 

\begin{figure}
\includegraphics[width=\columnwidth]{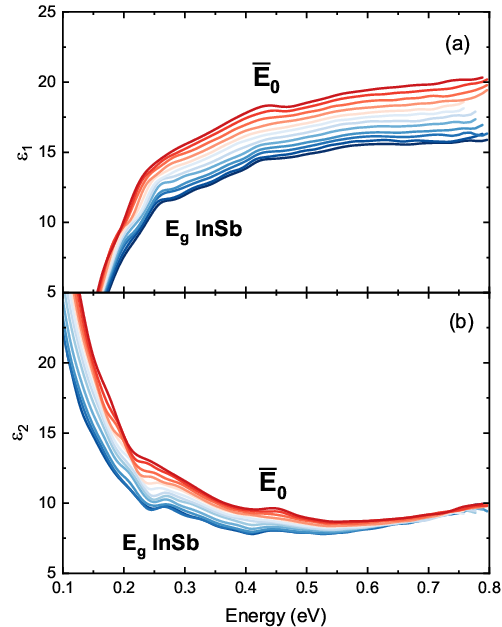}
\caption{Real (a) and imaginary (b) parts of the dielectric function $\epsilon$ of an Sb-doped $\alpha$-Sn layer (AE227) obtained from a basis spline fit. Same color scale as in Fig.\ \ref{psiDel}.}
\label{AE227-eps}
\end{figure}

\ifsec\section{Results}\fi \label{results}

The dielectric function for a nominally undoped $\alpha$-Sn layer on InSb (sample AE225) resulting from a B-spline fit is displayed in Fig.\ \ref{psiDel}. There are three artifacts in this dielectric function: First, there is a feature near 0.2 eV because of the incomplete removal of interference effects from the InSb substrate. (It is common to find such artifacts in the dielectric function of a layer near a critical point of the substrate. They are the result of improper assumptions in the data analysis, such as abrupt interfaces and homogeneous, plan-parallel layers.) Second, we did not consider surface overlayers, such as surface roughness, oxidation of $\alpha$-Sn, or (even in UHV at 10$^{-8}$ Torr) the condensation of residual gases on the surface. Ignoring such overlayers causes pseudo-absorption. This means that $\epsilon_2$ shown in Fig.\ \ref{psiDel} is only an upper limit. If an overlayer correction was performed, $\epsilon_2$ would be reduced. The impact of surface overlayers on $\epsilon_1$ is usually smaller than for $\epsilon_2$. Finally, the free carrier concentration of the substrate (which is not known for this particular sample) might influence the spectra at the lowest energies.

Despite these uncertainties, three results stand out in the dielectric function of $\alpha$-Sn shown in Fig.\ \ref{psiDel}: First, $\epsilon_1$ decreases and $\epsilon_2$ increases at the lowest energies. This divergence is more pronounced at higher temperatures. It arises from the combined free carrier response of the InSb substrate and the $\alpha$-Sn epilayer (which are difficult to separate). We will not discuss this Drude behavior of the epilayer further due to complications with the data analysis, because we do not currently have a good model for the temperature dependence of the Drude response of the InSb substrate. Second, we notice a steady increase of $\epsilon_2$ above 0.3~eV, while $\epsilon_1$ is flat in this region. The origin of this behavior is not certain. Part of it might be an artifact due to the presence of surface overlayers, which were ignored in the data analysis. It is suspicious that this linear increase becomes more pronounced at lower temperatures (where more ice might have formed on the sample). The opposite is expected for the intrinsic optical constants of $\alpha$-Sn. In the absence of artifacts, we expect that $\epsilon_2$ should increase with photon energy due to transitions from the $\Gamma_7^-$ inverted ``electron" valence band to the $\Gamma_8^+$ light ``hole" conduction band. A calculation based on parabolic bands from an 8$\times$8 $\vec{k}\cdot\vec{p}$-model finds $\epsilon_2$$<$1 at 0.8 eV for such transitions.\cite{Zo24} Nonparabolicity effects and remote bands increase this value, while the k-dependence of the optical dipole transition matrix element decreases it.\cite{Ka57} Two other possibilities include forbidden transitions with a k-linear optical dipole matrix element from the $\Gamma_7^+$ split-off hole band or from the $\Gamma_8^+$ valence band to the $\Gamma_8^+$ conduction band\cite{GeIv76} or an Urbach tail below the $E_1$ gap near 1.3 eV. 

Finally, the most prominent feature in the spectra (least impacted by uncertainties in the data analysis) is the $\bar{E}_0$ peak at 0.45 eV. The shape of this peak and its magnitude at 300~K have been analyzed elsewhere.\cite{Zo24} This peak is due to direct intravalence band transitions from the $\Gamma_7^-$ inverted ``electron" valence band to empty states (holes) in the $\Gamma_8^+$ heavy hole valence band. These transitions can only occur in the presence of holes in the $\Gamma_8^+$ band. Therefore, for an intrinsic sample, they are less pronounced at low temperature than at room temperature. The heavy hole density can also be tuned (increased or decreased) by doping the $\alpha$-Sn layer with In or Sb. 

The $E_0$ peak is therefore sample-dependent. An example of a dielectric function for an Sb-doped $\alpha$-Sn layer is shown in Fig.\ \ref{AE227-eps} (sample AE227). Because of a lower hole concentration in this n-type sample, the $\bar{E}_0$ peak is much weaker in this sample than in p-type or intrinsic $\alpha$-Sn layers. Other examples can be found in Ref.\ \citenum{CaZa18}. We note that the $\bar{E}_0$ peak in the ellipsometric angles and in the pseudodielectric function increases with thickness, but is independent of thickness in the dielectric function. 

\ifsec\section{Discussion}\fi \label{discussion} 

\begin{figure*}
\includegraphics[width=\textwidth]{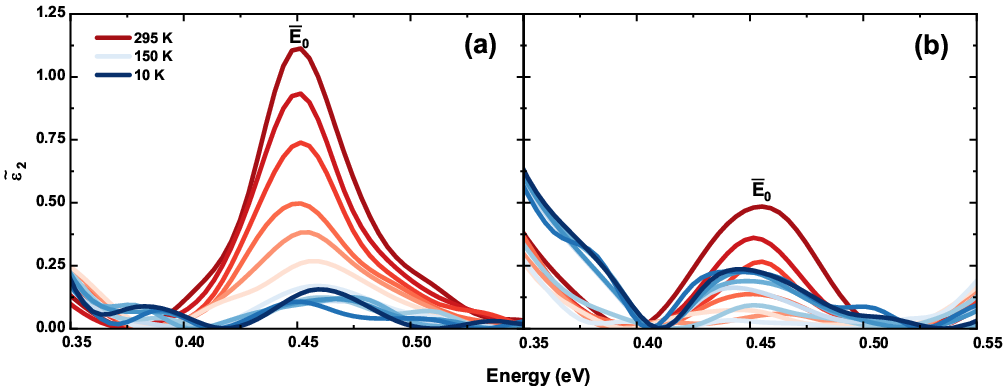}
\caption{Imaginary part of the dielectric function $\epsilon$ after subtracting the linear background for a nearly intrinsic (a) and Sb-doped (b) $\alpha$-Sn layer on InSb at different temperatures. Same color scale as in Fig.\ \ref{psiDel}.}
\label{eps-tilde}
\end{figure*}

The dielectric function $\epsilon$ of isotropic materials must meet many requirements due to physical constraints that originate from Maxwell's equations. For example, $\epsilon$ must be an analytic complex function. It must also satisfy causality, time-reversal symmetry, and the reality of the fields.\cite{St63} There are also other frequency-averaged conditions that restrain the optical constants. For example, the average refractive index (averaged over all frequencies) is equal to unity.\cite{Sm98} These conditions are often referred to as sum rules.

Since the complex dielectric function of a material describes the collective polarizability of charges under the influence of an electromagnetic wave with photon energy $\omega$, it is plausible that the magnitude of the dielectric function is related to the total charge density in the material. This is the so-called f-sum \hl{rule}\cite{AlDe72,Sm98,Ho70,Wo72}
\begin{equation}
\int_0^\infty\omega\epsilon_2\left(\omega\right)d\omega=\frac{\pi}{2}\omega_P^2=\frac{\pi}{2}\frac{ne^2\hbar^2}{\epsilon_0m_0},
\label{fsum}
\end{equation}
where $\omega_P$ is the \hl{unscreened} plasma energy and $n$ the carrier density. $\hbar$ is the reduced Planck's constant, $m_0$ is the free electron mass, and $\epsilon_0$ is the vacuum permittivity. The product $\omega\epsilon_2$ is proportional to the oscillator strength $f$, which explains the name of this sum rule.\cite{AlDe72} 
\hl{Integrating to very high energies is important, because this sum rule ignores the periodic potential of the crystal and treats all electrons as free.}\cite{St63} 
\hl{It also neglects the contribution of the nuclei because their mass is much larger than that of the electron.}\cite{St63}

\hl{As written,} Eq.\ (\ref{fsum}) should yield the valence electron density (e.g., four electrons per atom, about 1.8$\times$10$^{23}$ cm$^{-3}$ for a covalently bonded semiconductor such as Ge) if $m_0$ is taken as the free electron mass. Unfortunately, Eq.\ (\ref{fsum}) can rarely be used to obtain the valence electron density due to practical limitations: (1) Ellipso\-metry data for $\epsilon_2$ are rarely available above 10 eV. Therefore, typical experiments do not probe all charges.\cite{TsBa25} The limited spectral range requires fitting the available experimental data with a dispersion model followed by extrapolation to 50 eV, see Ref.\ \citenum{LiZh22}. For example, an integration over $\omega\epsilon_2$ for AlSb up to 5.8~eV yields only 2.4 electrons per atom.\cite{ZoLi89} (2) When integrating up to very large energies, the dielectric function might also include the contributions from transitions involving core electrons. This will overestimate the valence electron density. (3) Due to surface layers, such as roughness or native oxides, the experiment might underestimate the value of $\epsilon_2$ in the ultraviolet spectral region.

\hl{Our approach uses a finite-energy application of the sum rule, with well-defined lower and upper limits of the integral. This isolates the contribution of specific transitions that occur in this spectral range.}\cite{Sm98}
\hl{We also consider the effect of the periodic crystal potential by replacing the free electron mass with the effective mass of the electron and hole.}\cite{Wo72}
In our case, we are only interested in transitions from the $\Gamma_7^-$ ``electron" band to the $\Gamma_8^+$ heavy hole band, i.e., the contributions of heavy holes to the dielectric function. We therefore integrate from $E_a$=0.40 eV to $E_b$=0.55 eV, in the vicinity of the $\bar{E}_0$ peak. This results in the sum rule
\begin{equation}
\int_{E_a}^{E_b}\omega\tilde\epsilon_2\left(\omega\right)d\omega=\frac{\pi}{2}\frac{pe^2\hbar^2}{\epsilon_0m_0m_{hh}},
\label{fsumhh}
\end{equation}
where $\tilde\epsilon_2$ is the imaginary part of the dielectric function, from which the linear background (due to other types of transitions) has been subtracted. $p$ is the heavy hole density and $m_{hh}$ is the effective mass of the heavy holes. Figure \ref{eps-tilde} shows $\tilde\epsilon_2$ for one intrinsic and one Sb-doped $\alpha$-Sn layer. Since $\omega$$\approx$$\bar{E}_0$ over the narrow range of this integral, it is sufficient to pull $\bar{E}_0$ in front of the integral and integrate $\tilde\epsilon_2$ rather than the oscillator strength. 

\begin{figure}
\includegraphics[width=\columnwidth]{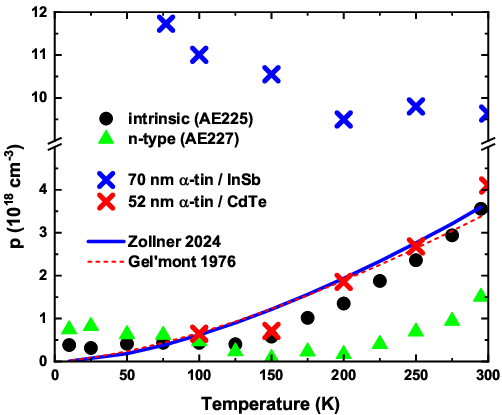}
\caption{Heavy hole density as a function of temperature for intrinsic and n-type $\alpha$-Sn layers determined from Eq.\ (\ref{fsumhh}) (symbols) in comparison to a calculation from Ref.\ \citenum{Zo24} based on degenerate Fermi-Dirac statistics (solid). Results from the literature\cite{CaZa18} for $\alpha$-Sn on InSb and CdTe are also shown.}
\label{phh}
\end{figure}

If the f-sum rule in Eq.\ (\ref{fsumhh}) is applied to the data in Fig.\ \ref{eps-tilde}, the obtained heavy hole concentrations lie between 0 (at 10 K) and 3$\times$10$^{18}$ cm$^{-3}$ (at room temperature) for the nearly intrinsic $\alpha$-Sn layer, in good agreement with calculations of the temperature-dependent heavy hole concentration based on degenerate Fermi-Dirac statistics with one hole band and two electron bands,\cite{Zo24} see Fig.\ \ref{phh}. Previous infrared ellipsometry measurements of an $\alpha$-Sn layer on CdTe (which is expected to be intrinsic) showed similar results.\cite{CaZa18} Hall measurements of an $\alpha$-Sn layer on CdTe also yielded heavy hole concentrations in the mid-10$^{18}$ cm$^{-3}$ range just below room temperature.\cite{HoMe89}

A second sample (AE227) yielded much lower heavy hole concentrations, reaching only 10$^{18}$ cm$^{-3}$ at 300 K. We conclude that this layer, grown on an Sb-rich substrate surface, is doped with electrons from Sb donors, which reduce the thermal population of heavy holes at low temperature. An $\alpha$-Sn layer on InSb grown by a different group\cite{CaZa18} shows a significantly higher heavy hole concentration, even at low temperatures. We conclude that this sample has additional holes due to In acceptors, even at temperatures below 100 K. All data are summarized and compared with Fermi-Dirac statistics in Fig.\ \ref{phh}. 

\ifsec\section{Summary}\fi  \label{summary} 

In summary, Hall measurements of $\alpha$-Sn layers grown by Hartmut H\"ochst's group in Wisconsin using molecular beam epitaxy showed many years ago\cite{HoMe89} that $\alpha$-Sn layers grown on In-doped CdTe substrates result in p-type $\alpha$-Sn layers while those grown on undoped CdTe tend to be intrinsic or n-type. Growth of $\alpha$-Sn on InSb substrates has also resulted in p-type $\alpha$-Sn layers.\cite{CaZa18,BaDu14,MiSo87} It is difficult to reproduce such Hall measurements on modern samples, because producing Hall bars on thin $\alpha$-Sn layers\cite{EnDe24} and Hall effect data analysis\cite{LaEw71} tend to be cumbersome. Therefore, we have demonstrated the use of a non-destructive all-optical technique, which produces similar hole concentration results as Hall measurements. Our method only determines the hole concentration. Its accuracy is about 5$\times$10$^{17}$ cm$^{-3}$. It is not affected by the complicated conduction band structure caused by the near degeneracy of the $\Gamma$- and $L$-bands.\cite{Ew68}

Second, we were able to show that not only the choice of the substrate\cite{HoMe89,CaZa18} can modulate the electron and hole doping levels, but also the surface preparation of the InSb substrate before growth. An In-rich InSb substrate surface produces p-type or nearly intrinsic $\alpha$-Sn layers, while Sb-rich InSb substrate surfaces can result in n-type layers. 

\begin{acknowledgments}
This material is based upon work supported by the National Science Foundation under award number DMR-2423992,
by the Air Force Office of Scientific Research under award number FA9550-24-1-0061, 
and by the Department of Energy, National Nuclear Security Administration under award number DE-NA0004103. The growth at UCSB was supported by the Army Research Laboratory (Grants No. W911NF-21-2-0140 and No. W911NF-23-2-0031). CAA gratefully acknowledges support from the J.\ A.\ Woollam Foundation. JH was supported by a Fulbright-Masaryk Award. JRL received DOD SCALE support from the US Department of the Navy under award number W52P1J-22-9-3009. 
We are grateful to Arnold Kiefer for stimulating discussions. 
\end{acknowledgments}


\section*{Conflict of interest}

The authors have no conflicts to disclose.

\section*{Data availability}

The data that support the findings of this study are available from the corresponding author upon reasonable request

\end{document}